\documentclass[pra,aps,12pt]{revtex4-2}
\usepackage{amssymb}
\usepackage{graphicx}

\begin{document}

\title{Macroscopic degeneracy of ground state in the frustrated Heisenberg
diamond chain}
\author{D.~V.~Dmitriev}
\email{dmitriev@deom.chph.ras.ru}
\author{V.~Ya.~Krivnov}
\affiliation{Institute of Biochemical Physics of RAS, Kosygin str.
4, 119334, Moscow, Russia.}
\date{}

\begin{abstract}
The spin-$\frac{1}{2}$ Heisenberg diamond chain with ferro- and
antiferromagnetic exchange interactions is studied. The phase
boundary in the parametric space of these interactions is
determined, where the transition between the ferromagnetic and
other (singlet or ferrimagnetic) ground state phases occurs. On
this phase boundary there is a dispersionless (flat) energy band
in the one-magnon spectrum and these states can be represented as
localized magnons in the trapping cells between neighboring
diamonds. The ground state consists of the localized magnons and
special magnon complexes and it is macroscopically degenerate. A
remarkable feature of the model is the existence, on a certain
part of the phase boundary, of two- and three-magnon localized
states, forming flat bands in two- and three-magnon spectrum. All
these localized states also belong to the ground state manifold,
which turns out to be exactly the same as for the system of
independent spins-$\frac{3}{2}$. This implies a large macroscopic
degeneracy of the ground state $4^n$ ($n$ is number of diamonds in
the chain) and a high residual entropy per spin
$s_{0}=\frac{2}{3}\ln 2$.
\end{abstract}

\maketitle

\section{Introduction}

The low-dimensional quantum magnets on geometrically frustrated
lattices have been extensively studied during the last years
\cite{diep,mila}. There is a peculiar class of frustrated quantum
magnets in which the ground state is macroscopically degenerate.
Systems with macroscopic degeneracy have attracted great interest
and are usually described by models with flat bands
\cite{Flach,flat}. Flat-band in the one-magnon spectrum of the
frustrated Heisenberg spin system means that the magnons are
localized in a small portion of the lattice (trapping cells). The
localization of the magnons arises due to a destructive quantum
interference, caused by frustration. There are highly frustrated
antiferromagnetic spin systems which support a completely
dispersionless magnon band \cite{flat, mak, zhit, *zhit2, shulen},
so that the excitations in this band are localized states. The
localization of one-magnon states is a base for the construction
of multi-magnon states, because a state consisting of independent
(non-overlapping) localized magnons is an exact eigenstate. Such
systems include, for example, the delta-chain, the kagome lattice,
kagome-like chains, the Tasaki lattice etc. Their important
feature is the triangular geometry of antiferromagnetic bonds. The
ground state of such systems at the saturation magnetic field
consists of independent localized magnons and it is
macroscopically degenerate \cite{flat}. The ground state and
low-temperature properties for the antiferromagnetic Heisenberg
models with flat band have been actively studied over the last
decades. It was shown that flat band physics may lead to new
interesting phenomena such as the residual entropy at the
saturation magnetic field, the zero-temperature magnetization
plateau and the magnetization jump, an extra low-temperature peak
in the specific heat etc. \cite{hon, schmidt, honecker,shulen,
Derzhko}. Such systems have enhanced magnetocaloric effect and are
promising for efficient low-temperature cooling
\cite{schmidt,honecker,zhitomir,Derzhko}.

\begin{figure}[tbp]
\includegraphics[width=5in,angle=0]{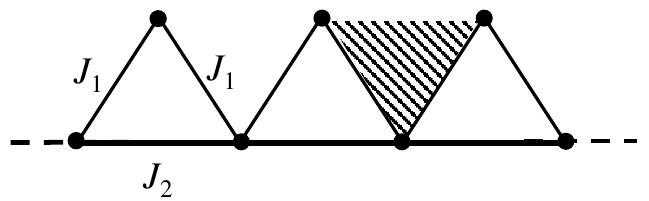}
\caption{Delta chain spin model. Shaded area indicates the trapping valley.}
\label{Fig_delta_chain}
\end{figure}

Recently it was found that the localized magnon states can exist
in a certain frustrated spin system with competing ferro- (F) and
antiferromagnetic (AF) interactions. An example of such model is
spin-$\frac{1}{2}$ Heisenberg delta-chain with the ferromagnetic
$J_{1}<0$ and the antiferromagnetic $J_{2}>0$ interactions (see
Fig.\ref{Fig_delta_chain}). For $J_{2}=-\frac{J_{1}}{2}$ the
localized states exist in this model. However, in contrast to the
AF-AF delta-chain, the localized magnon states of the F-AF chain
are exact ground states \textit{at zero magnetic field}. Besides,
the ground state manifold contains the special states with
overlapping magnons (localized multi-magnon complexes). Thus, the
ground state degeneracy in this model is even higher than the one
for the AF-AF delta chain. The properties of such F-AF delta chain
have been studied in Refs.\cite{KDNDR,DKRS,*DKRS2,*DKRS3}.

\begin{figure}[tbp]
\includegraphics[width=5in,angle=0]{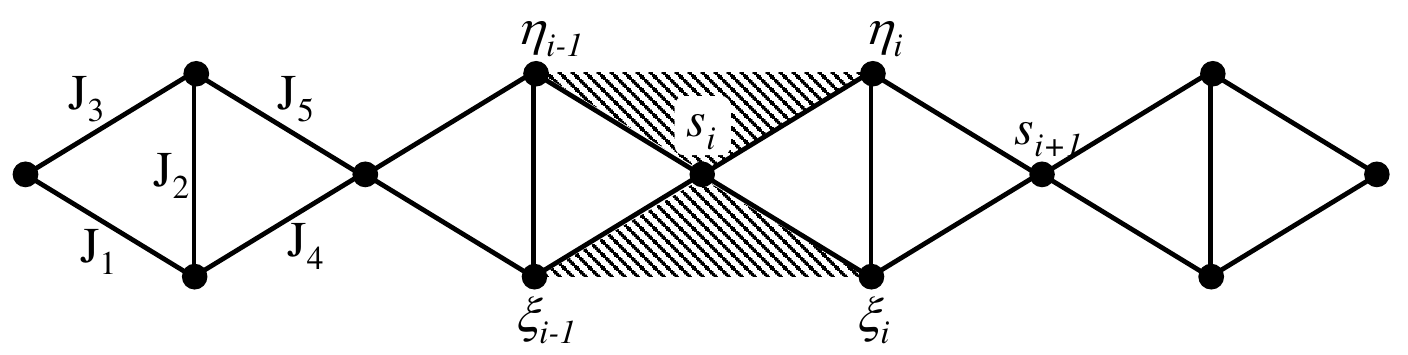}
\caption{Diamond chain spin model. Shaded area indicates the trapping cell.}
\label{Fig_rhomb_chain}
\end{figure}

In this paper we consider one more example of the model, in which
the localized magnons lead to the macroscopic degeneracy of the
ground state. This is the spin-$\frac{1}{2}$ Heisenberg diamond
chain, shown in Fig.\ref{Fig_rhomb_chain}. Generally, this model
is one of the examples where the interplay of quantum effects and
frustration leads to a wide variety of ground state phases; it has
attracted a lot of attention in both experimental and theoretical
studies. It was proposed as a minimal theoretical model of several
copper-based compounds as azurite $Cu_{3}(CO_{3})_{2}(OH)_{2}$
\cite{azur} and alumoklyuchevskite
$K_{3}Cu_{3}AlO_{2}(SO_{4})_{4}$ \cite{alum}. Besides, the diamond
model is also interesting in its own right. The ground state phase
diagram of the diamond chain has been extensively studied in many
works \cite{Takano,Hone,Strec,Sak,Sakai,Hu,Krupn,Gu,Okam,Morita}
and it was shown that the diamond model exhibits various quantum
phases depending on exchange interactions. In particular, the
$1/3$ magnetization plateau was determined for certain ratios
between exchange interactions \cite{Kitazawa,Kab,Okam,Gu,Morita}.

In this work we will study the spin-$\frac{1}{2}$ Heisenberg
diamond chain with F and AF competing interactions. We will show
that under special relations between exchange interactions,
describing the boundary between the ferromagnetic and other
(singlet or ferrimagnetic) phases, the ground state is
macroscopically degenerate.

The paper is organized as follows. In Sec.II the model Hamiltonian
with generally different exchange interactions is presented and
the conditions for the interactions in which the ground state has
macroscopic degeneracy are determined. It will be shown that,
under certain conditions, along with the localized one-magnon
states there are also localized two- and three-magnon states. We
present four types of the diamond chain with different ground
state degeneracy. In the first type only one-magnon localized
states exist. In the second one there are one-, two- and three-
magnons in the trapping cell and this case includes the diamond
chain with equivalent opposite sides of a diamond (we name it as
the `distorted' diamond chain). The third model is the ideal
(symmetric) diamond chain and in the fourth type of the model
one-magnon states are localized on diagonals of diamonds. The
ground state degeneracy for these four types of diamond chain
models are studied in Subsecs. IIA, IIB, IIC and IID. In
concluding Section we give a summary of our results.

\section{Diamond chain at the critical point}

The Hamiltonian of the spin $s=\frac{1}{2}$ Heisenberg diamond
chain can be represented as a sum of local Hamiltonians
\begin{equation}
\hat{H}=\sum_{i=1}^{n}\hat{H}_{i}  \label{H}
\end{equation}%
and $\hat{H}_{i}$ is the Hamiltonian of $i$-th diamond%
\begin{equation}
\hat{H}_{i}=J_{1}\mathbf{s}_{i}\cdot \mathbf{\xi }_{i}+J_{2}\mathbf{\xi }%
_{i}\cdot\mathbf{\eta }_{i}+J_{3}\mathbf{s}_{i}\cdot\mathbf{\eta }_{i}+J_{4}%
\mathbf{s}_{i+1}\cdot\mathbf{\xi }_{i}+J_{5}\mathbf{s}_{i+1}\cdot\mathbf{%
\eta }_{i}-\frac{J_{0}}{4}  \label{h}
\end{equation}%
where $\mathbf{s}_{i}$, $\mathbf{\xi }_{i}$, $\mathbf{\eta }_{i}$
are spin operators of $i$-th diamond, $n$ is total number of
diamonds and $J_{0}=J_{1}+J_{2}+J_{3}+J_{4}+J_{5}$. In general,
there are five different exchange interactions $J_{i}$ as shown in
Fig.\ref{Fig_rhomb_chain}. The constant in Eq.(\ref{h}) is chosen
so that the energy of the state with the maximal spin of diamond
($S=2$) is zero.

It is known \cite{KDNDR} that the macroscopic ground state
degeneracy in frustrated spin systems with competing (F) and (AF)
interactions arises if the local Hamiltonian has several
degenerate ground states, one of which is the state with the
maximal spin. Four spins of diamond form one quintet ($S=2 $),
three triplets ($S=1$) and two singlets ($S=0$). The condition
that the quintet and one of three triplets are degenerate with
zero energy is
\begin{equation}
J_{2}=-\frac{J_{1}J_{3}}{J_{1}+J_{3}}-\frac{J_{4}J_{5}}{J_{5}+J_{4}}
\label{rel1}
\end{equation}

The conditions that these eight states (5 states of quintet and 3
states of triplet) are the ground states with $E_{i}=0$ of
$\hat{H}_{i}$ and all other eigenvalues $E_{i}>0$ are
\begin{equation}
J_{1}+J_{3}<0,\qquad J_{4}+J_{5}<0  \label{rel2}
\end{equation}

Relations (\ref{rel1}) and (\ref{rel2}) determine the range of
interaction values corresponding to the transition point between
the ferromagnetic and other (singlet or ferrimagnetic) ground
state phases of model (\ref{H}). For simplicity we chose the
interaction $J_{2}$ as a parameter defining the transition point.
Eq.(\ref{rel1}) gives the critical value of this parameter. For
$J_{2}$ less than the value given by Eq.(\ref{rel1}), the ground
state of model (\ref{H}) is ferromagnetic. Further we assume that
both relations (\ref{rel1}) and (\ref{rel2}) are satisfied.

Because the neighboring local Hamiltonians $\hat{H}_{i}$ do not
commute with each other, the following inequality for the lowest
eigenvalue $E_{0}$ of $\hat{H}$ is valid:%
\begin{equation}
E_{0}\geq \sum E_{i}=0  \label{ineq}
\end{equation}

The energy of the ferromagnetic state of $\hat{H}$ with maximal
total spin $S_{\max }=\frac{3n}{2}$ is zero. Therefore, the
inequality in (\ref{ineq}) turns into an equality and the ground
state energy is zero.

The relations (\ref{rel1}) and (\ref{rel2}) can also be obtained
alternatively from the condition that a one-magnon excitation band
with the spin $S^{z}=S_{\max }^{z}-1$ is dispersionless (flat
band) with zero energy. The dispersionless one-magnon states
correspond to localized states which can be chosen as
\begin{equation}
\hat{\varphi}_{1i}\left\vert F\right\rangle =(\sigma
_{i-1}^{-}+s_{i}^{-}+\tau _{i}^{-})\left\vert F\right\rangle \text{ \qquad }%
i=1,2,\ldots n  \label{f1}
\end{equation}%
where $\left\vert F\right\rangle $ is the ferromagnetic state with
all spins up, $s_{i}^{-}$ are lowering spin operators and new
convenient operators $\sigma _{i}^{-}$, $\tau _{i}^{-}$ relate to
lowering spin operators $\eta _{i}^{-},\xi _{i}^{-}$ as follows:
\begin{eqnarray}
\sigma _{i}^{-} &=&\frac{J_{4}+J_{5}}{J_{1}J_{5}-J_{3}J_{4}}(J_{1}\eta
_{i}^{-}-J_{3}\xi _{i}^{-})  \label{sigma} \\
\tau _{i}^{-} &=&-\frac{J_{1}+J_{3}}{J_{1}J_{5}-J_{3}J_{4}}(J_{4}\eta
_{i}^{-}-J_{5}\xi _{i}^{-})  \label{tau}
\end{eqnarray}%
so that $\sigma _{i}^{-}+\tau _{i}^{-}=\eta _{i}^{-}+\xi _{i}^{-}$.

The states $\hat{\varphi}_{1i}\left\vert F\right\rangle $ are
localized in the trapping cells between neighboring diamonds as
shown in Fig.\ref{Fig_rhomb_chain} by a shaded area. All $n$
states (\ref{f1}) are the ground states and they are linearly
independent, because each $\hat{\varphi}_{1i}$ includes the
corresponding operator $s_{i}^{-}$, which is not shared with other
functions $\hat{\varphi}_{1j}$.

There are three one-magnon bands of model (\ref{H}), one of them
is flat, and the energies of the other two bands are given by
\begin{equation}
2E(q)=-J_{0}\pm \sqrt{J_{0}^{2}-3J_{2}(J_{0}-J_{2})-3(J_{1}+J_{4})(J_{3}-J_{5})-2(J_{3}J_{5}+J_{1}J_{4})(1-\cos q)%
}  \label{Ek}
\end{equation}

It can be shown that the energy $E(q)>0$ if both conditions (\ref{rel1}) and
(\ref{rel2}) are satisfied.

As follows from Eqs.(\ref{sigma}),(\ref{tau}), there is one special case in
which the denominator in these equations goes to zero:
\begin{equation}
\frac{J_{3}}{J_{1}}=\frac{J_{5}}{J_{4}}  \label{c2}
\end{equation}

In this special case the one-magnon states (\ref{f1}) are localized on
diagonals of diamonds and can be written as
\begin{equation}
\hat{\phi}_{i}\left\vert F\right\rangle =(J_{1}\eta _{i}^{-}-J_{3}\xi
_{i}^{-})\left\vert F\right\rangle \text{ \qquad }i=1,2,\ldots n
\label{diag}
\end{equation}

Therefore, we name this special case as `diagonal state case', the
properties of the model for this special case will be studied separately in
Sec.IId.

A remarkable feature of model (\ref{H}) is the fact that along with the
presence of one-magnon localized states, under certain condition, there are
exact two- and three-magnon localized states with zero energy. These
localized states form flat-bands in two- and three-magnon spectrum. Omitting
all intermediate calculations we give the final expressions for two- and
three-magnon states, localized in the $i$-th trapping cell (Fig.\ref%
{Fig_rhomb_chain}):
\begin{eqnarray}
\hat{\varphi}_{2i}\left\vert F\right\rangle &=&(\sigma
_{i-1}^{-}s_{i}^{-}+s_{i}^{-}\tau _{i}^{-}+\sigma _{i-1}^{-}\tau
_{i}^{-})\left\vert F\right\rangle  \label{f2} \\
\hat{\varphi}_{3i}\left\vert F\right\rangle &=&\sigma
_{i-1}^{-}s_{i}^{-}\tau _{i}^{-}\left\vert F\right\rangle  \label{f3}
\end{eqnarray}

It can be checked that Eqs.(\ref{f2}),(\ref{f3}) give exact two- and
three-magnon eigenstates of the Hamiltonian (\ref{H}) with zero energy if
the exchange interactions satisfy the relation
\begin{equation}
\frac{J_{3}}{J_{1}}=\frac{J_{4}}{J_{5}}  \label{c1}
\end{equation}

Then, two- and three-magnon states (\ref{f2}) and (\ref{f3}) also
belong to the ground state manifold. It is worth noting that when
the relation (\ref{c1}) is valid, the ground state degeneracy of
each local Hamiltonian $\hat{H}_{i}$ increases. In this case, one
singlet state is added to 8 degenerate ground states with $S=2$
and $S=1$.

Relation (\ref{c1}) is satisfied automatically for the important
case of the `distorted' diamond chain, in which $J_{3}=J_{4}$ and
$J_{1}=J_{5}$ (but $J_{1}\neq J_{3}$).

There is one more special case, when both relations (\ref{c2}) and
(\ref{c1}) are satisfied simultaneously. In this case
$J_{1}=J_{3}=J_{4}=$ $J_{5}=-1$ and $J_{2}=1$ (ideal diamond
chain) and localized one-magnon functions are diagonal singlets
$(\eta _{i}^{-}-\xi _{i}^{-})\left\vert F\right\rangle $. As it
will be shown below, the ground state degeneracy is maximal in
this case.

\begin{figure}[tbp]
\includegraphics[width=5in,angle=0]{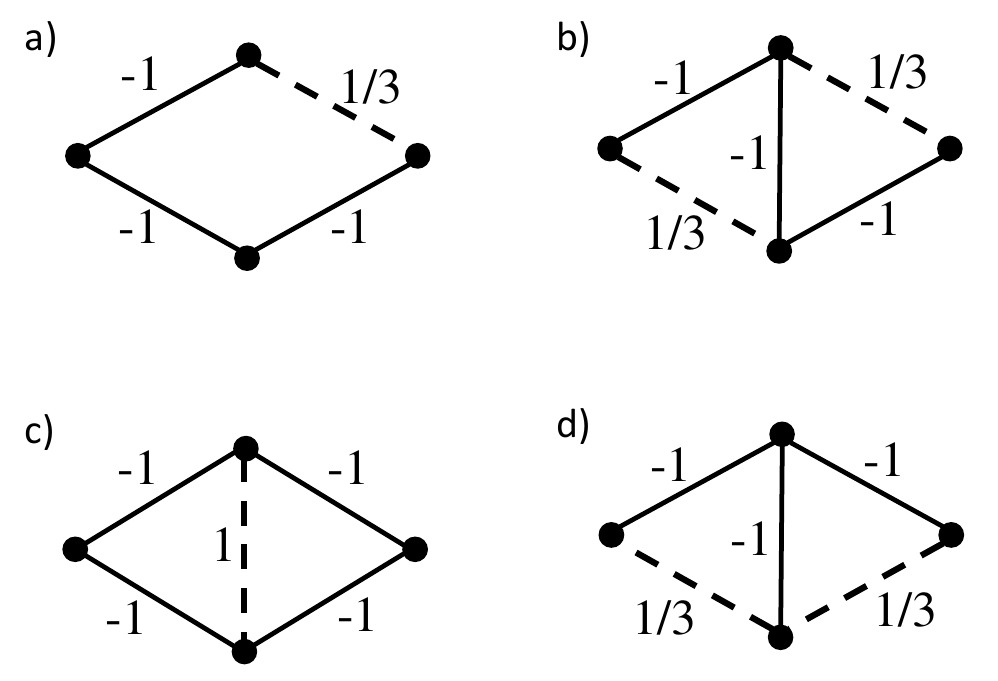}
\caption{Examples of four different types of diamond spin chains
with macroscopic ground state degeneracy. The values of the
exchange integrals are written close to the corresponding bonds.
For clarity, the ferromagnetic bonds (all with $J=-1$) are shown
by solid lines, antiferromagnetic bonds - by dashed lines.}
\label{Fig_all_cases}
\end{figure}

Thus, we have determined the necessary conditions (\ref{rel1}) and
(\ref{rel2}) for the macroscopic ground state degeneracy of the
diamond chain model (\ref{H}). Based on the above analysis we can
distinguish four different types of the diamond chain with the
macroscopic ground state degeneracy, depending on whether the
relations (\ref{c2}) and (\ref{c1}) are satisfied or not. Examples
of four types of diamonds are shown in Fig.\ref{Fig_all_cases}. We
will name these types as follows: the `general case' is the one
when none of the relations (\ref{c2}) and (\ref{c1}) are fulfilled
(Fig.\ref{Fig_all_cases}a); the `distorted diamond case' is
realized when the relation (\ref{c1}) is satisfied, but
Eq.(\ref{c2}) is not (Fig.\ref{Fig_all_cases}b); the `ideal
diamond case' is when both relations are fulfilled
(Fig.\ref{Fig_all_cases}c); and the `diagonal diamond case' is
when the relation (\ref{c2}) is satisfied, but Eq.(\ref{c1}) is
not (Fig.\ref{Fig_all_cases}d).

\begin{figure}[tbp]
\includegraphics[width=5in,angle=0]{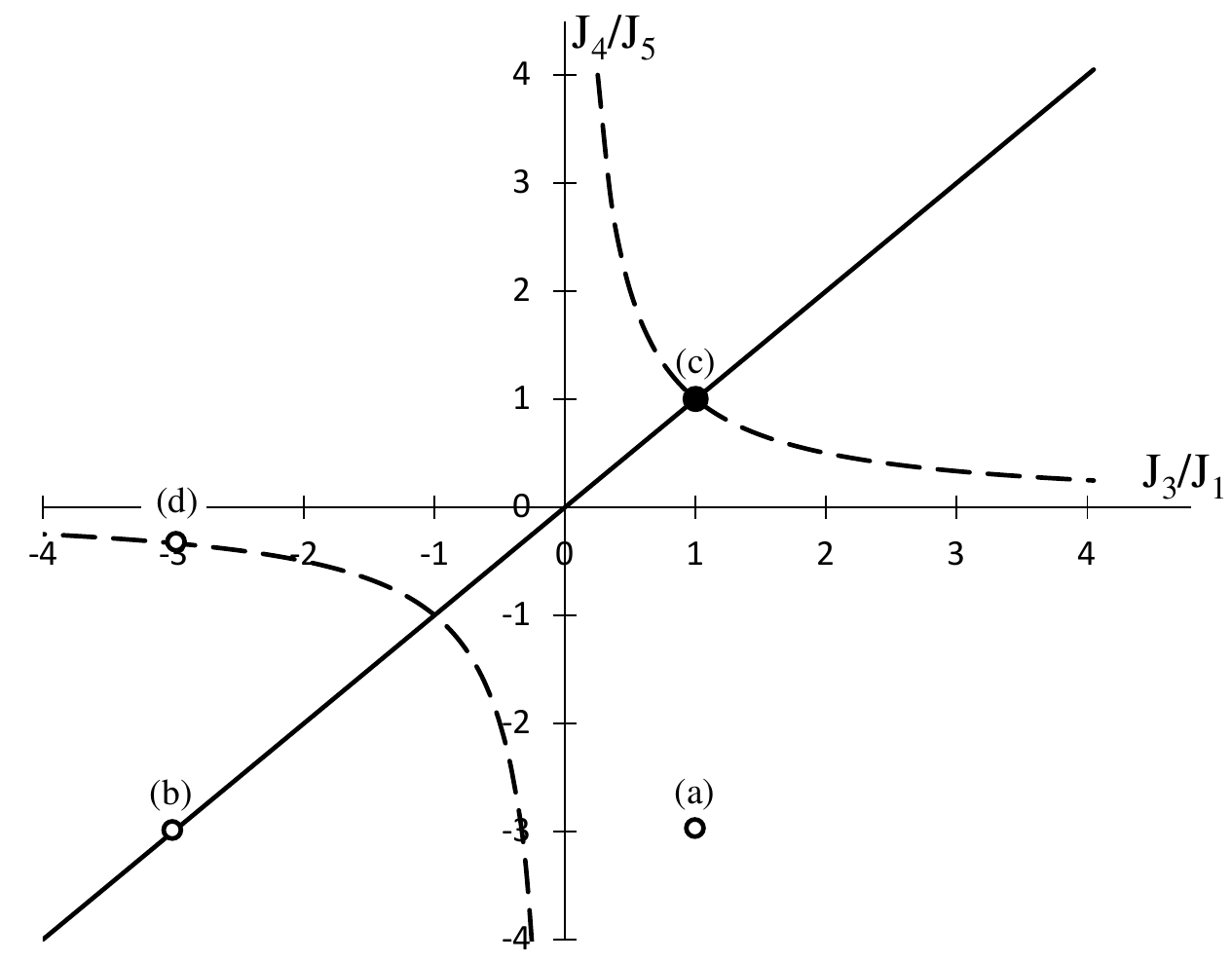}
\caption{Phase diagram of spin diamond chain (\protect\ref{H})
with diagonal bond interaction $J_2$ given by
Eq.(\protect\ref{rel1}). The special cases are shown by solid line
(`distorted diamond case'), dashed line (`diagonal state case')
and by solid circle (`ideal diamond case'). Thick open circles
denote four particular cases of the diamond model shown in
Fig.\ref{Fig_all_cases} and they are signed with the corresponding
letters a, b, c and d.} \label{Fig_phase}
\end{figure}

As it follows from the relations (\ref{c2}) and (\ref{c1}), it is
convenient to plot the phase diagram of the model (\ref{H}) in the
$J_{3}/J_{1}$ and $J_{5}/J_{4}$ axes (see Fig.\ref{Fig_phase}). In
Fig.\ref{Fig_phase} the `diagonal state case' is shown by a dashed
line, the distorted diamond case by a solid line and the ideal
case by a solid circle. The entire plane, except for these special
cases, corresponds to the general case. However, in order to
satisfy the conditions (\ref{rel2}), there is a prescription on
how to choose the signs of interactions $J_{i}$ corresponding to a
definite point on the phase diagram Fig.\ref{Fig_phase}. If the
point is in the region $J_{3}/J_{1}>0$, then the condition
$J_{1}+J_{3}<0$ leads uniquely to the conclusion that both $J_{3}$
and $J_{1}$ are ferromagnetic ($J_{3}$, $J_{1}<0$). If
$-1<J_{3}/J_{1}<0$, then $J_{3}>0$ and $J_{1}<0$. And, finally, if
$J_{3}/J_{1}<-1$, then $J_{3}<0$ and $J_{1}>0$. Similar conditions
are applied for the correct choice of signs of interactions $J_{4}
$ and $J_{5}$ in different regions of the phase diagram. To
illustrate, four particular cases of the diamond model shown in
Fig.\ref{Fig_all_cases}, are represented by thick open circles in
Fig.\ref{Fig_phase} and signed with the corresponding letters: a,
b, c and d.

\subsection{General case}

Let us first consider the case when both relations (\ref{c2}) and
(\ref{c1}) are not satisfied. The example of the diamond of this
type is shown in Fig.\ref{Fig_all_cases}a, where the exchange
integrals are chosen so that the diagonal interaction $J_{2}$ is
zero. In the general case there are only one-magnon localized
states with zero energy as in the F-AF delta-chain at the critical
value of the frustration parameter. In the latter case the
one-magnon state with zero energy is localized in a valley between
neighboring triangles (see Fig.\ref{Fig_delta_chain}). Any state
consisting of independent (isolated) magnons has zero energy and
belongs to the ground state manifold. As it is shown in
\cite{KDNDR} for the F-AF delta-chain, in addition to the
independent localized states there are also specifically
overlapping magnons in the ground states. For the general case of
the diamond chain the problem of counting of these states is
completely similar to the one for the F-AF delta-chain, which was
solved in \cite{KDNDR}. Rewriting these results to the case of the
diamond chain with periodic boundary conditions (PBC) we find that
the number of ground states $W(n,k)$ in the spin sector
$S^{z}=S_{\max }-k$ is
\begin{eqnarray}
W(n,k) &=&C_{n}^{k}\qquad 0\leq k\leq n/2 \\
W(n,k) &=&C_{n}^{n/2}\qquad n/2+1\leq k\leq 3n/2
\end{eqnarray}%
where $C_{n}^{k}=\frac{n!}{k!(n-k)!}$ are binomial coefficients.

The total number of the ground states $W(n)$ is%
\begin{equation}
W(n)=2^{n}+(n+4)C_{n}^{n/2}
\end{equation}

According to this equation the ground state degeneracy is
exponentially large. It leads to the finite residual entropy per
spin (total number of spins is $N=3n$)
\begin{equation}
s_{0}=\frac{\ln W(n)}{N}  \label{entropy}
\end{equation}
and $s_{0}=\frac{1}{3}\ln 2$ at $n\gg 1$.

\begin{figure}[tbp]
\includegraphics[width=5in,angle=0]{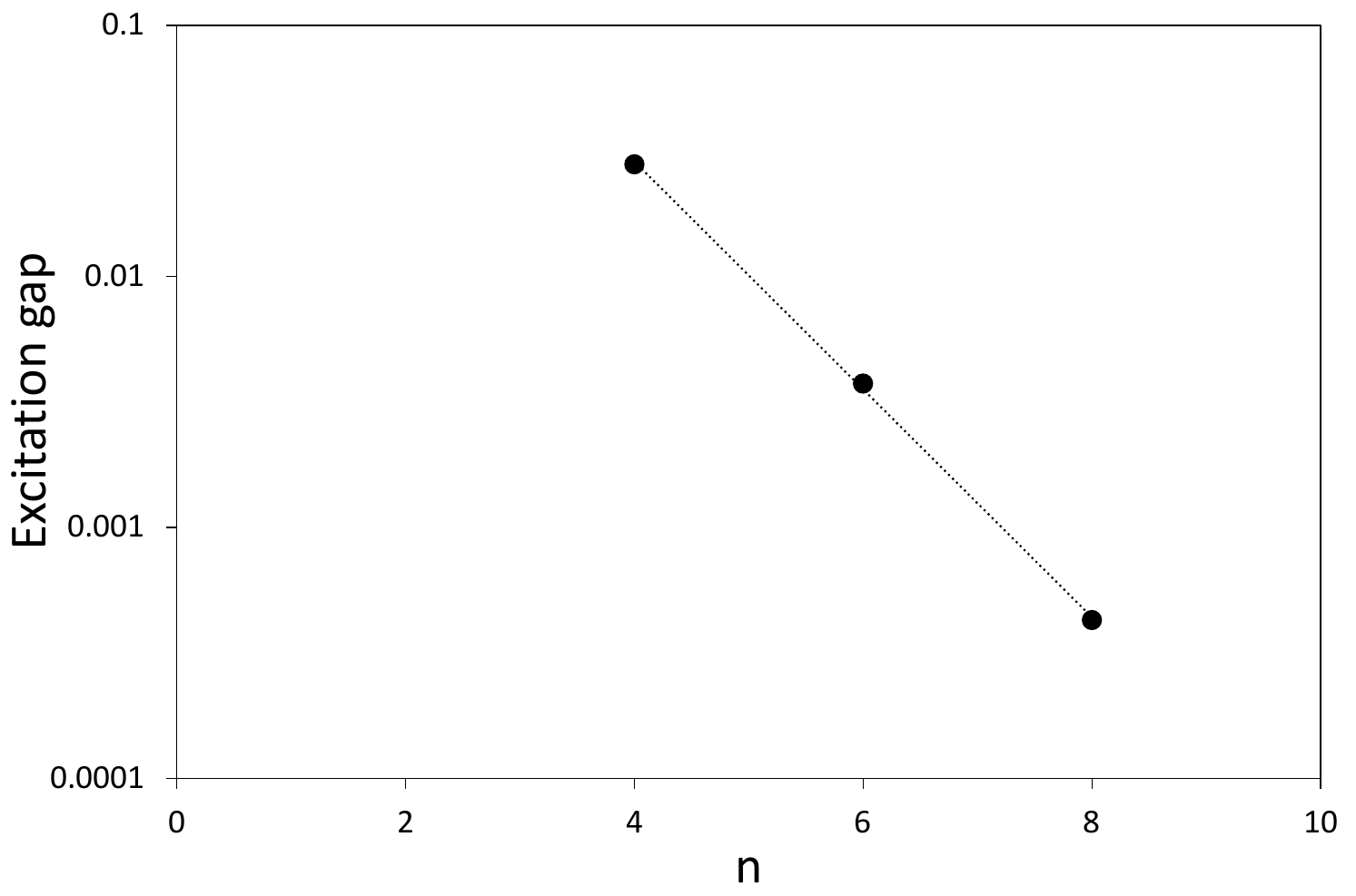}
\caption{Dependence of the energy gap (logarithmic scale) on the
system size for the general case (Fig.\ref{Fig_all_cases}a),
calculated numerically for $N=12,18,24$ (exact diagonalization and
Lanczos algorithm).} \label{Fig_gap_general}
\end{figure}

The properties of the excitation spectrum of the diamond chain in
the general case are similar to that for the F-AF delta-chain.
Numerical calculations show that the gap in the spectrum is
exponentially small at $n\gg 1$ as in the F-AF delta-chain (see
Fig.\ref{Fig_gap_general}).

\begin{figure}[tbp]
\includegraphics[width=5in,angle=0]{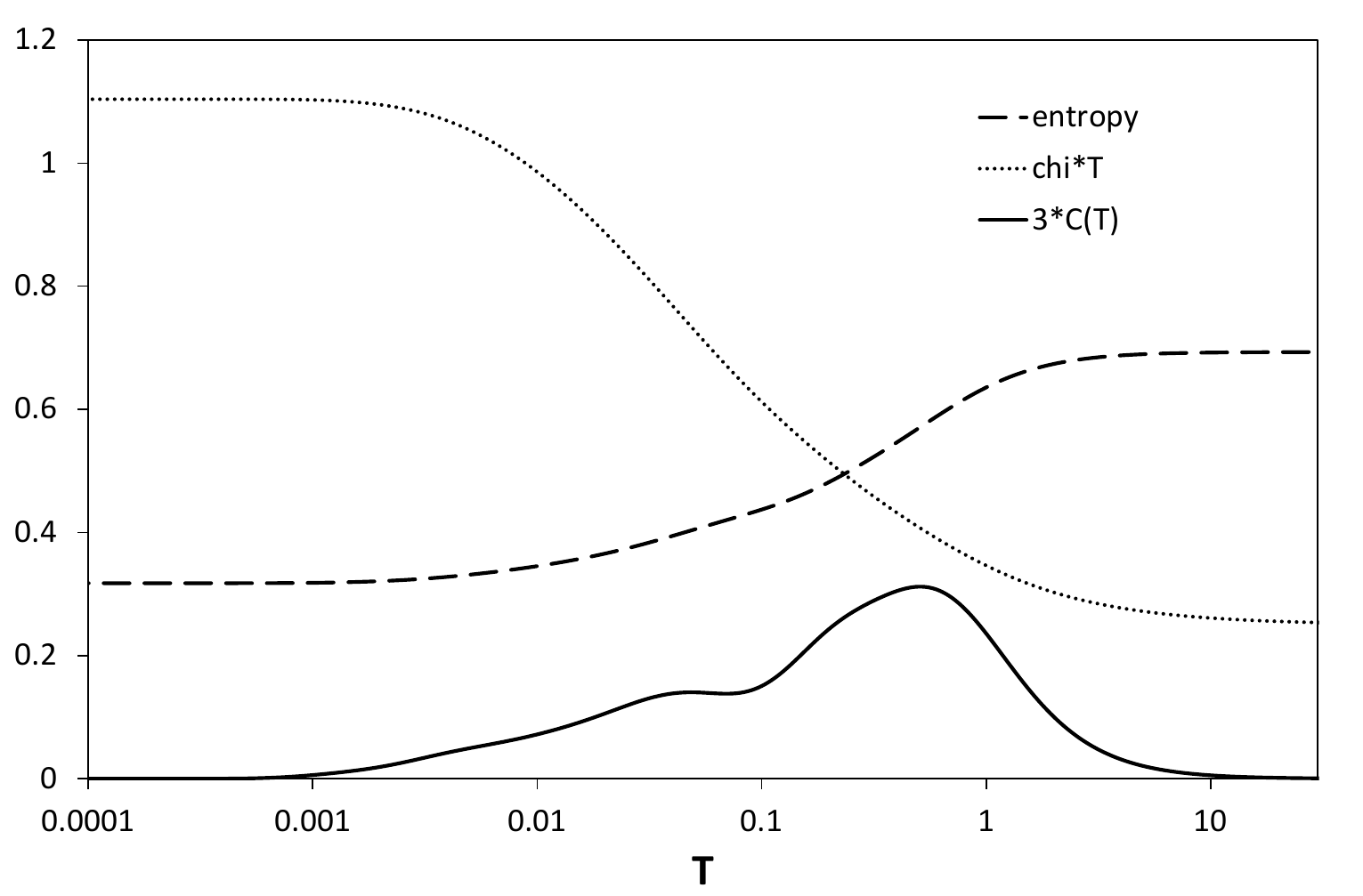}
\caption{Temperature dependencies of the specific heat (solid
line), entropy (dashed line) and the product of magnetic
susceptibility and temperature, $\chi T$, (dotted line) for the
general case (Fig.\ref{Fig_all_cases}a), calculated numerically
for $N=18$ (exact diagonalization). For better visibility the
specific heat is multiplied by a factor of 3, $3C(T)$.}
\label{Fig_T_general}
\end{figure}

The low-temperature thermodynamic properties in this case are
determined by these low-lying excitations and they lead to the
specific behavior of the thermodynamic quantities at low
temperature. For example, $C(T)$ has low-temperature maximum and a
long low-temperature tail, as shown in Fig.\ref{Fig_T_general}.
The entropy in Fig.\ref{Fig_T_general} tends to the finite value
in accordance with Eq.(\ref{entropy}) at $T\to 0$ and increases to
the value $s_{0}=\ln 2$ at $T\to\infty$.

The behavior of the magnetization and the uniform susceptibility
of the diamond chain (in the general case) is similar to the
behavior of the F-AF delta-chain, and this point was discussed in
detail in \cite{KDNDR}. We give here a brief summary of the
results \cite{KDNDR}.

Since there are exponentially low excited states, the contribution
of excited states to magnetization and susceptibility can not be
neglected even at very low temperatures. Nevertheless, it is
useful to consider the magnetization given only by the
contribution of the ground state manifold. This magnetization
$m_{0}$ is a function of variable $x=\frac{h}{T}$. At
$x\gg\frac{1}{N}$ the magnetization $m_{0}$ for the general case
of diamond chain is
\begin{equation}
m=\frac{M}{N}=\frac{1}{2}-\frac{1}{3(1+\exp (h/T))}  \label{m(h)}
\end{equation}

\begin{figure}[tbp]
\includegraphics[width=5in,angle=0]{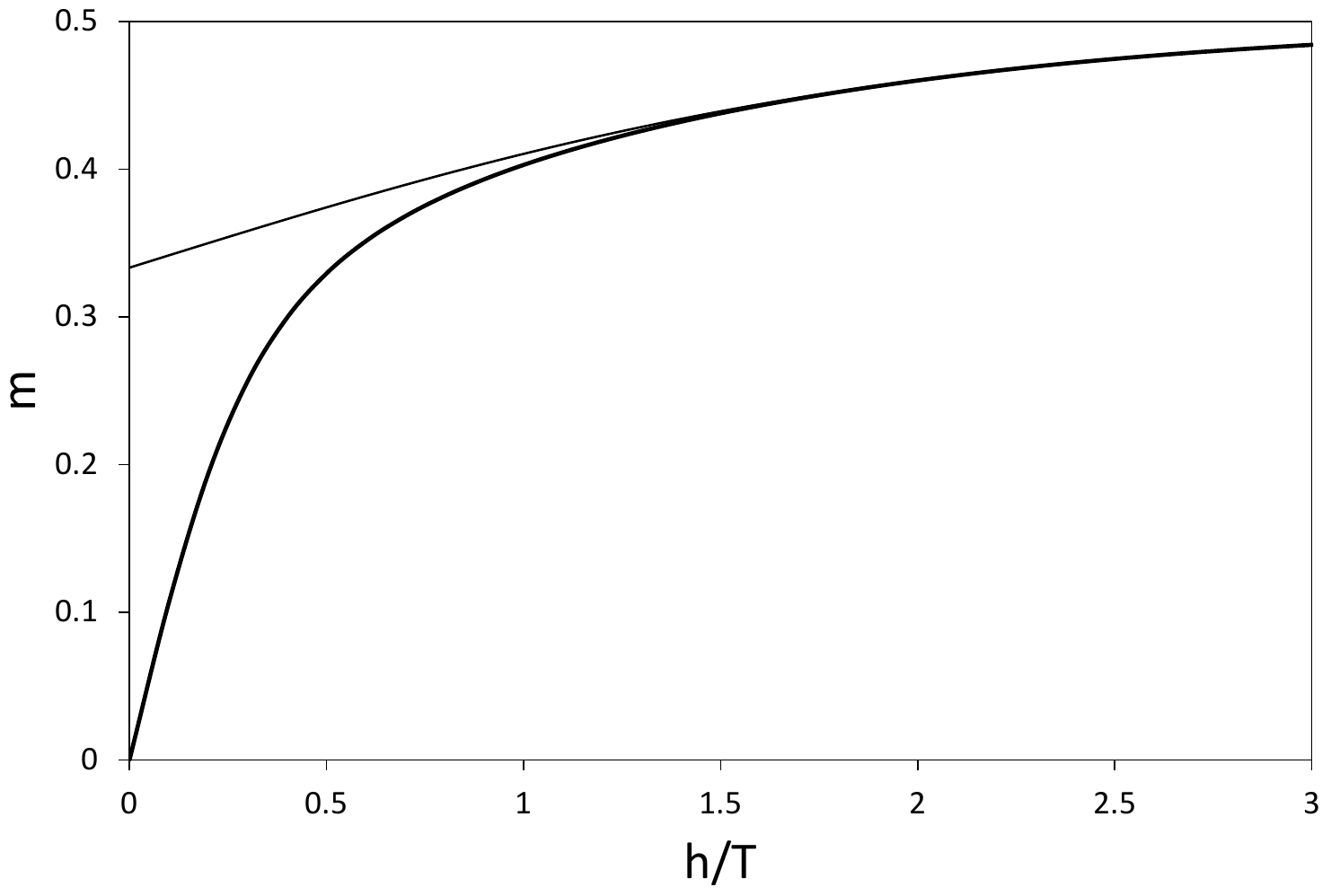}
\caption{Magnetization curve as a function of $h/T$ for the
general case (Fig.\ref{Fig_all_cases}a), calculated numerically
for $N=18$ and $T=0.001$ (thick solid line) and given by
Eq.(\ref{m(h)}) (thin solid line).} \label{Fig_Mh_general}
\end{figure}

The magnetization curve (\ref{m(h)}) is shown in
Fig.\ref{Fig_Mh_general} together with the magnetization curve
calculated numerically for $N=18$ and $T=0.001$. As one can see in
Fig.\ref{Fig_Mh_general}, the calculated curve coincides with the
curve (\ref{m(h)}) for $h/T>1$, but the limit $h/T\to 0$ is
different. The reason of this discrepancy is related to the
singular behavior of magnetization and susceptibility in low
magnetic fields.

The numerically calculated magnetization is zero at $h=0$ and
increases linearly with increasing magnetic field, $m=\chi h$. The
magnetic susceptibility $\chi$ depends on the temperature and the
system size. An analysis similar to the one given in \cite{KDNDR}
shows that in the thermodynamic limit the magnetic susceptibility
diverges at $T\to 0$ as $\chi\sim T^{-\alpha}$ with the exponent
$\alpha\approx 1.25$. Thus, the true magnetization curve in the
thermodynamic limit, $m(h)$, starts at $m(0)=0$, then increases
linearly with a very steep slope,
\begin{equation}
m \sim \frac{h}{T^{\alpha}}
\end{equation}
so that it approaches the value $m=1/3$ at $h_c\sim T^{\alpha}$.
For $h>h_c$ the true magnetization is described by
Eq.(\ref{m(h)}).

When the magnetization curve is plotted as a function of $h/T$, as
shown in Fig.\ref{Fig_Mh_general}, the low-field region where the
true magnetization curve deviates from Eq.(\ref{m(h)}) is limited
by $h_c/T\sim T^{\alpha-1}$, so that this region vanishes in the
limit $T\to 0$. This implies that the magnetization at
$\frac{h}{T}\to 0$ tends to $\frac{1}{3}$. Therefore, the diamond
chain in the general case is magnetically ordered at the
transition point at $T=0$ and, therefore, the magnetization
undergoes a jump from $m=\frac{1}{2}$ in the ferromagnetic phase
to $m=\frac{1}{3}$ at the transition point.

\subsection{`Distorted' diamond chain}

Let us consider the `distorted' diamond chain in which the
trapping cell can contain localized one-, two- and three-magnon
states. In this case condition (\ref{c1}) is satisfied but
(\ref{c2}) is not. Further in this subsection we focus on the most
interesting from the experimental point of view case $J_{3}=J_{4}$
and $J_{1}=J_{5}$ (but assuming $J_{1}\neq J_{3}$). A particular
case of such choice corresponds to the distorted diamond chain
shown in Fig.\ref{Fig_all_cases}b. In fact, this is a generic case
for the the `distorted' diamond chain, so that the results for the
case $J_{3}\neq J_{4}$, $J_{1}\neq J_{5}$ are qualitatively
similar.

All possible states consisting of independent (isolated) magnons have zero
energy and belong to the ground states. However, in similarity to the
previous case such states do not exhaust all possible ones with zero energy.
For example, definite combinations of overlapping magnons are also the exact
ground states. For the F-AF delta chain with open boundary condition (OBC)
such multi-magnon states can be chosen as products
\begin{equation}
\Phi _{1}(m_{1})\Phi _{1}(m_{2})\ldots \Phi _{1}(m_{k})\left\vert
F\right\rangle \qquad 1\leq m_{1}<m_{2}<\ldots <m_{k}\leq n-1  \label{prod1}
\end{equation}%
where%
\begin{equation}
\Phi _{1}(m)=2S^{-}(m)+\eta _{m+1}^{-}
\end{equation}%
and $S^{-}(m)$ is the total lowering spin operator of all spins
for the first $m$ triangles and $\eta _{m+1}^{-}$ is the lowering
operator of $(m+1)$-th apical spin.

As it was shown in \cite{KDNDR}, the product (\ref{prod1}) is the
exact eigenfunction of each triangle Hamiltonian including $m$-th
and $(m+1)$-th triangles.

For the diamond chain the analog of the operator $\Phi _{1}(m)$
has the form
\begin{equation}
\Phi _{1}(m)=S^{-}(m)+\tau _{m+1}^{-}
\label{S1}
\end{equation}%
where $\tau _{m+1}^{-}$ is provided in Eq.(\ref{tau}) and
\begin{equation}
S^{-}(m)=s_{m+1}^{-}+\sum_{i=1}^{m}(s_{i}^{-}+\xi _{i}^{-}+\eta _{i}^{-})
\end{equation}

For the diamond chain with two- and three-magnons localized in the
same trapping cell this approach can be extended to include
functions similar to $\Phi _{1}(m)$ for two- and three-magnon
states. These functions have the forms
\begin{eqnarray}
\Phi _{2}(m) &=&S^{-}(m)[\frac{1}{2}S^{-}(m)+\tau _{m+1}^{-}]  \label{S2} \\
\Phi _{3}(m) &=&S^{2-}(m)[\frac{1}{3}S^{-}(m)+\tau _{m+1}^{-}]  \label{S3}
\end{eqnarray}

It can be checked that the functions $\Phi _{\sigma }(m)\left\vert
F\right\rangle $ ($\sigma =1,2,3$) are exact eigenfunctions of all
local Hamiltonians with zero energy. Then we can construct the
following products for the diamond model with (OBC)
\begin{equation}
\Phi _{\sigma _{1}}(m_{1})\Phi _{\sigma _{2}}(m_{2})\ldots \Phi _{\sigma
_{k}}(m_{k})\left\vert F\right\rangle \text{ \ \ }1\leq m_{1}<m_{2}<\ldots
<m_{k}\leq n-1
\end{equation}%
where $\sigma _{i}=1,2,3$.

These products are the zero-energy eigenfunctions of each local
Hamiltonian and they form the ground states. We can interpret
these products as the states of the system of $n$ non-interacting
spins $\frac{3}{2}$. For example, let us consider the product
\begin{equation}
\Phi _{1}(m_{1})\Phi _{3}(m_{2})\Phi _{2}(m_{3})\Phi _{2}(m_{4})\left\vert
F\right\rangle
\end{equation}

This product can be identified as the state of the system with
spin $s=\frac{3}{2}$ consisting of $(n-4)$ spins with
$s^{z}=\frac{3}{2}$, one spin with $s^{z}=\frac{1}{2}$, two spins
with $s^{z}=-\frac{1}{2}$ and one spin with $s^{z}=-\frac{3}{2}$.
Generally, the number of ground states of the diamond model in the
spin sector $S^{z}$ coincides with that for the system of
non-interacting spins $\frac{3}{2}$.

Further, we consider the case of the periodic boundary conditions
(PBC). Any state consisting of isolated magnons
\begin{equation}
\varphi _{\sigma _{1}i_{1}}\varphi _{\sigma _{2}i_{2}}\varphi _{\sigma
_{3}i_{3}}\ldots \varphi _{\sigma _{k}i_{k}}\left\vert F\right\rangle
,\qquad i_{l}>i_{l-1}+1
\end{equation}%
is an exact ground state. But these non-overlapping magnon states
do not exhaust all possible ground states. For example, let us
consider the overlapping magnon state
\begin{equation}
\Omega _{\sigma _{1}}(i)\varphi _{\sigma _{2}i}\left\vert F\right\rangle
\label{overlap}
\end{equation}%
where%
\begin{eqnarray}
\Omega _{1}(i) &=&\sigma _{i-1}^{-}+S^{-}(i-1,i)+\tau _{i+1}^{-} \\
\Omega _{2}(i) &=&S^{-}(i-1,i)[\sigma _{i-1}^{-}+\frac{1}{2}%
S^{-}(i-1,i)+\tau _{i+1}^{-}] \\
\Omega _{3}(i) &=&S^{2-}(i-1,i)[\sigma _{i-1}^{-}+\frac{1}{3}%
S^{-}(i-1,i)+\tau _{i+1}^{-}]
\end{eqnarray}%
and $S^{-}(i-1,i)$ is the total lowering operator of ($i-1)$-th
and $i$-th diamonds. It can be checked that the functions of the
type (\ref{overlap}) are the eigenstates with zero energy of the
local Hamiltonians $\hat{H}_{i-1} $, $\hat{H}_{i}$, and of all the
others.

We conjecture that in similarity to the OBC case the problem of
counting the number of ground states is reduced to the one for a
system of $n$ non-interacting spins $\frac{3}{2}$. Then, the
number of ground states $W(n,k)$ in the spin sector $S^{z}=S_{\max
}-k$ is determined by equations
\begin{eqnarray}
k &=&n_{1}+2n_{2}+3n_{3}  \nonumber \\
l &=&n_{1}+n_{2}+n_{3}  \label{comb}
\end{eqnarray}%
where $(n-l),n_{1},n_{2},n_{3}$ are numbers of spins
$s=\frac{3}{2}$ with $s_{z}=\frac{3}{2}$, $\frac{1}{2}$,
$-\frac{1}{2}$, $-\frac{3}{2}$, correspondingly. The solution of a
combinatorial problem (\ref{comb}) gives the number of ground
states $W(n,k)$ in form
\begin{equation}
W(n,k)=\sum_{l}\sum_{m}C_{n}^{l}C_{l}^{k-l-m}C_{k-l-m}^{m}  \label{W}
\end{equation}

Using (\ref{W}) we can count up the total ground state degeneracy
$W(n)$
\begin{equation}
W(n)=\sum_{k}W(n,k)=4^{n}  \label{W1}
\end{equation}

Results (\ref{W}) and (\ref{W1}) have been confirmed by ED
calculation of finite diamond chains. The residual entropy per
spin in accordance with Eq.(\ref{entropy}) is
\begin{equation}
s_{0}=\frac{2}{3}\ln 2 \label{s0_distorted}
\end{equation}

\begin{figure}[tbp]
\includegraphics[width=5in,angle=0]{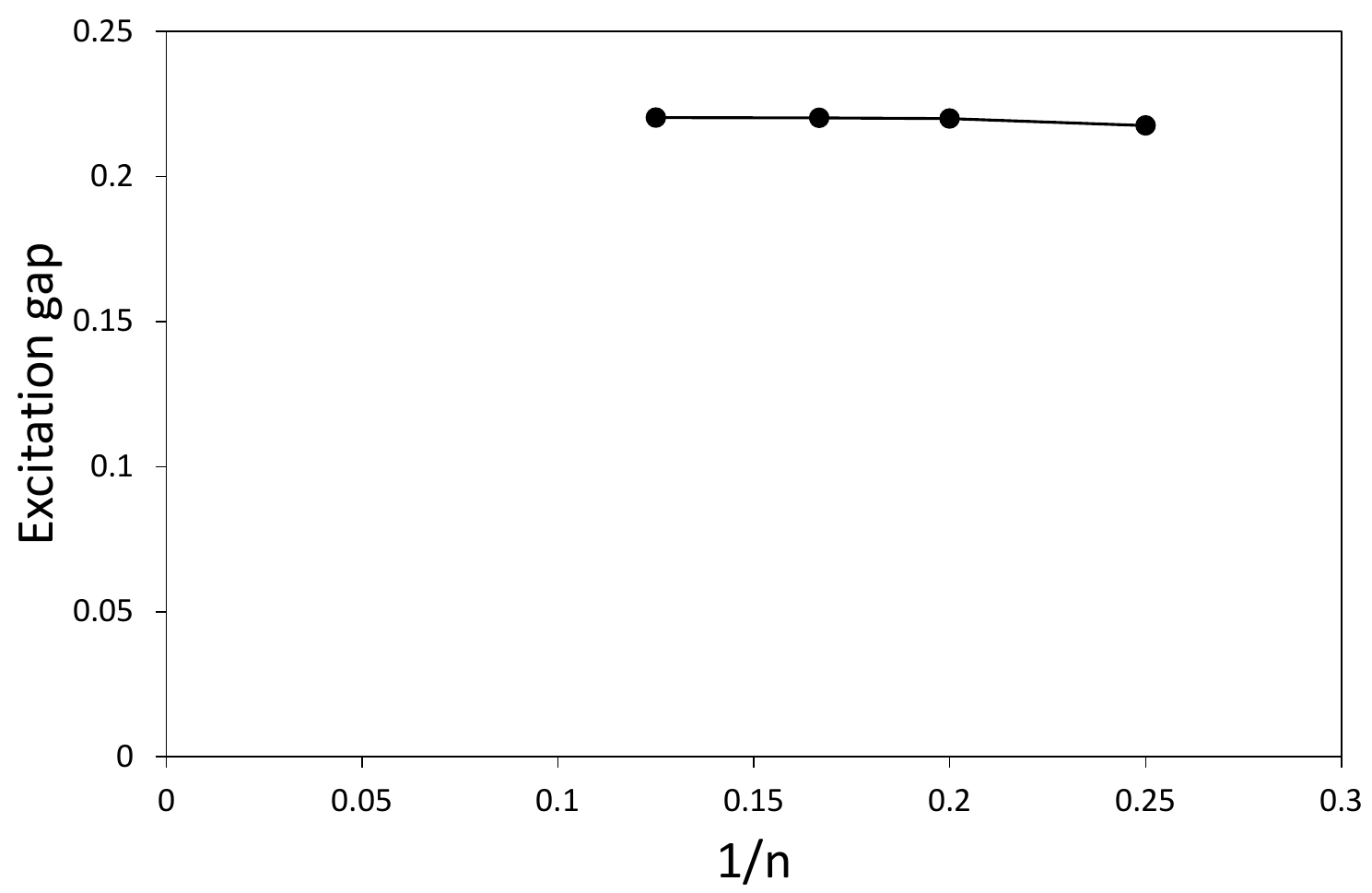}
\caption{Dependence of the energy gap on the system size for
distorted diamond case shown in Fig.\ref{Fig_all_cases}b,
calculated numerically for $N=12,15,18,24$.}
\label{Fig_gap_distorted}
\end{figure}

The excitation spectrum of the distorted diamond chain at the
transition point is gapped. The gap in the one-magnon spectrum
$\Delta E_{1}$, given by Eq.(\ref{Ek}) for the diamond in
Fig.\ref{Fig_all_cases}b, is $\Delta E_{1}=1/3$. Numerical
calculations show that the minimal gap is lower, but still finite
$\Delta E\simeq 0.22$ (see Fig.\ref{Fig_gap_distorted}).
Therefore, for $T<\Delta E$ the system properties are determined
by the contribution of the ground state manifold or, equivalently,
it is given by the one for the system of $n$ non-interacting spins
$\frac{3}{2}$.

\begin{figure}[tbp]
\includegraphics[width=5in,angle=0]{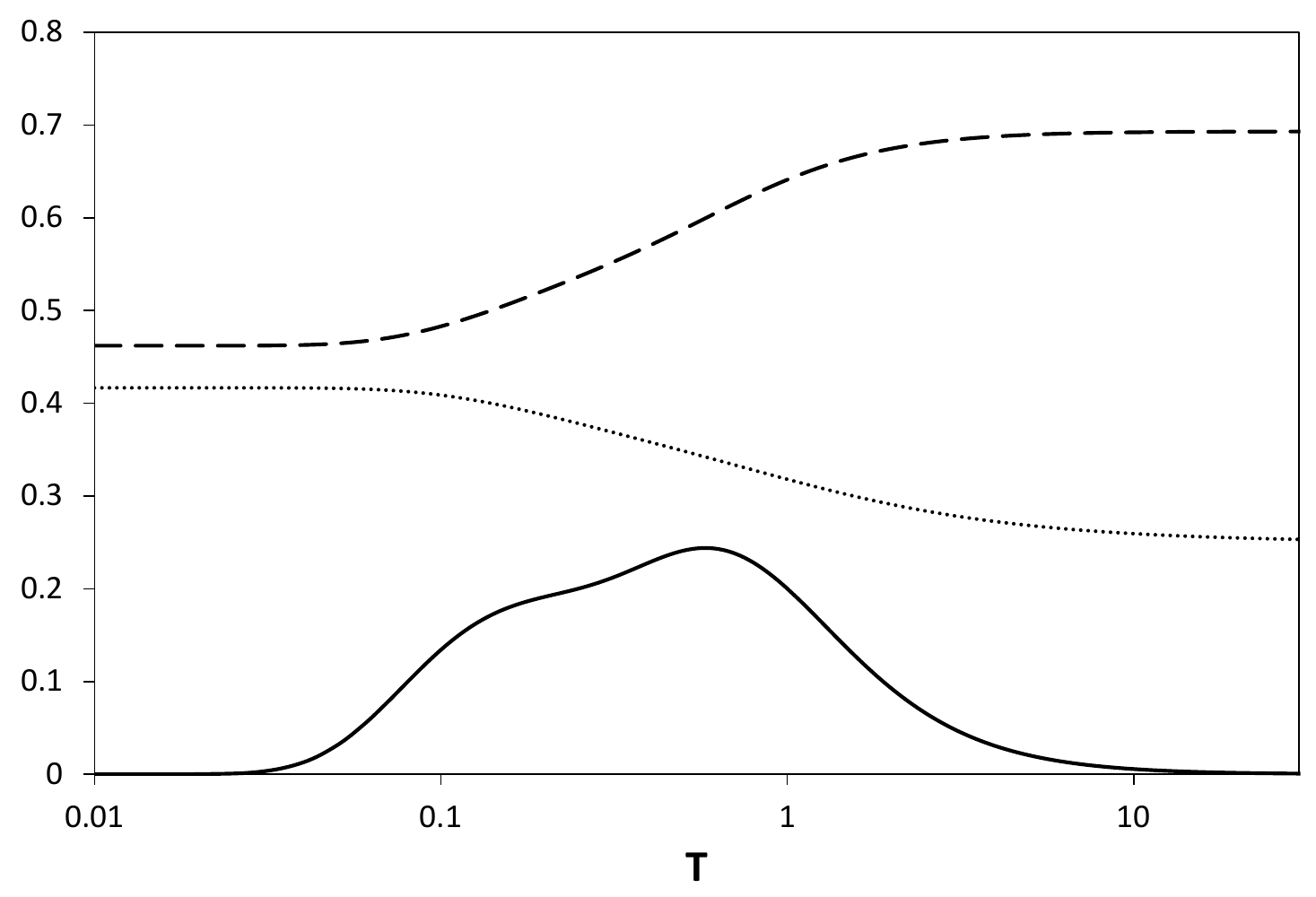}
\caption{Temperature dependencies of the specific heat (solid
line), entropy (dashed line) and the product of magnetic
susceptibility and temperature, $\chi T$, (dotted line) for the
distorted diamond case shown in Fig.\ref{Fig_all_cases}b,
calculated numerically for $N=18$. For better visibility the
specific heat is multiplied by a factor of 3, $3C(T)$.}
\label{Fig_T_distorted}
\end{figure}

The thermodynamic properties of the distorted diamond chain shown
in Fig.\ref{Fig_all_cases}b are demonstrated in
Fig.\ref{Fig_T_distorted}. As it is seen in
Fig.\ref{Fig_T_distorted}, the entropy is equal to its residual
value (\ref{s0_distorted}) at low temperature $T<\Delta E$ and
tends to its high temperature limiting value $\ln 2$. The specific
heat $C(T)$ has one maximum with a wide left shoulder. The
susceptibility per spin for $T<\Delta E$ behaves as for the system
of non-interacting spins $\frac{3}{2}$: $\chi =\frac{5}{12T}$. For
high temperature the susceptibility tends to $\chi =\frac{1}{4T}$.

\begin{figure}[tbp]
\includegraphics[width=5in,angle=0]{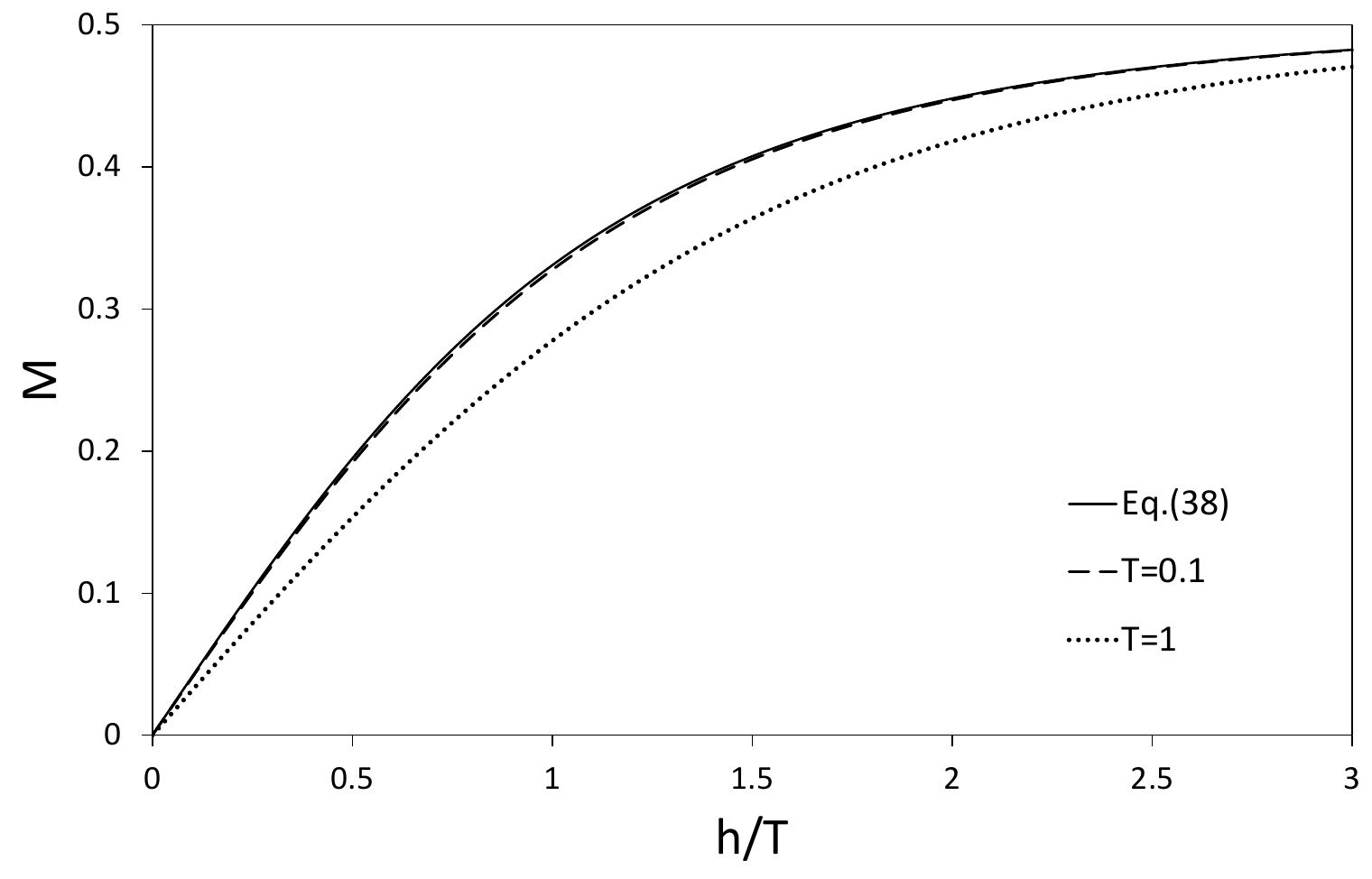}
\caption{Magnetization curve as a function of $h/T$ for distorted
diamond case shown in Fig.\ref{Fig_all_cases}b, calculated
numerically for $N=18$ and $T=0.1$ (dashed line) and $T=1$ (dotted
line). Magnetization curve (\ref{mh_distorted}) is shown by solid
line.} \label{Fig_Mh_distorted}
\end{figure}

The magnetization per spin of the system of non-interacting spins
$\frac{3}{2}$ is given by
\begin{equation}
m=\frac {3\sinh(\frac{3h}{2T})+\sinh(\frac{h}{2T})}
{6\cosh(\frac{3h}{2T})+6\cosh(\frac{h}{2T})} \label{mh_distorted}
\end{equation}
and is shown in Fig.\ref{Fig_Mh_distorted} by solid line. As one
can see in Fig.\ref{Fig_Mh_distorted}, the magnetization curve for
$T=0.1$, which is lower than the energy gap, perfectly coincides
with Eq.(\ref{mh_distorted}). But the magnetization curve for
relatively high temperature $T=1$ substantially deviates from
Eq.(\ref{mh_distorted}).

\subsection{Ideal diamond chain}

For the ideal diamond chain the interactions $J_{i}$ are
$J_{1}=J_{3}=J_{4}=J_{5}=-1$ and $J_{2}=1$ (see
Fig.\ref{Fig_all_cases}c). In this special case there is a local
conservation of the composite spin $\mathbf{L}_{i}=(\mathbf{\xi
}_{i}+\mathbf{\eta }_{i})$ on the vertical diagonals of the
diamonds. The composite spin is a conserved quantity with quantum
spin number $L_{i}=0$ or $L_{i}=1$. Then the model reduces to the
spin chain with alternating spins $s=\frac{1}{2}$ and composite
spins $L$, and it is described by the Hamiltonian
\begin{equation}
\hat{H}=-\sum [\mathbf{L}_{i}\cdot (\mathbf{s}_{i}+\mathbf{s}_{i+1})-1]+%
\frac{1}{2}(\mathbf{L}_{i}^{2}-2)  \label{L}
\end{equation}%
where $\mathbf{L}_{i}$ and $\mathbf{s}_{i}$ are composite spin and
spin $s=\frac{1}{2}$, respectively, on $i$-th diamond. The
constants in Eq.(\ref{L}) are chosen so that the ground state
energy of Hamiltonian (\ref{L}) is zero.

All ground states of (\ref{L}) can be enumerated as follows.
First, we choose a definite configuration of $k$ singlets $L=0$
located in diamonds $\{i_{1},i_{2},\ldots i_{k}\}$. The number of
such configurations with $k$ singlets is $C_{n}^{k}$. We note that
the singlet state on the diagonal of the diamond is an exact state
for this diamond, independent of the spin states on the left and
the right sites of it. Effectively, each singlet $L_{i}=0$ cuts
the chain and creates an open boundary at this place. Therefore,
for the chosen configuration of singlets $\{i_{1},i_{2},\ldots
i_{k}\}$, we have $k$ sections of open chains, located between
spins with $L=0$. The ground state of each of these open chains is
the ferromagnetic state with all possible projections $S^{z}$,
which can be written as
\begin{equation}
\psi \left( S_{ij}^{z}\right) =\left\vert
S_{max}(i,j),S_{ij}^{z}\right\rangle  \label{Ff}
\end{equation}%
where the total spin of the ferromagnetic state of the open
diamond chain between the singlets on diagonals of $i$-th and
$j$-th diamonds is $S_{max}(i,j)=\frac{3}{2}(j-i)-1$ and the spin
projection can be $S_{ij}^{z}=-S_{max},\ldots S_{max}$.
Consequently, the number of multiplets of the ferromagnetic state
between the singlets on $i$-th and $j$-th diamonds is
$Q_{ij}=3(j-i)-1$.

Thus, the set of ground state wave functions for the configuration of
singlets $\{i_{1},i_{2},\ldots i_{k}\}$ can be written as
\begin{equation}
\Psi \left( S_{i_{1}i_{2}}^{z},S_{i_{2}i_{3}}^{z},\ldots
S_{i_{k}i_{1}}^{z}\right) =\left\vert s_{i_{1}}\right\rangle \psi \left(
S_{i_{1}i_{2}}^{z}\right) \left\vert s_{i_{2}}\right\rangle \psi \left(
S_{i_{2}i_{3}}^{z}\right) \ldots \left\vert s_{i_{k}}\right\rangle \psi
\left( S_{i_{k}i_{1}}^{z}\right)  \label{GL}
\end{equation}%
where $\left\vert s_{i}\right\rangle $ means the singlet state on
diagonal of $i$-th diamond. Therefore, the set of wave functions
(\ref{GL}) contains $Q_{i_{1}i_{2}}\cdot Q_{i_{2}i_{3}}\cdot
\ldots \cdot Q_{i_{k}i_{1}}$ states. To calculate the total number
of ground states one should sum up all such products for all
possible singlet configurations $\{i_{1},i_{2},\ldots i_{k}\}$
with different $k$. The solution of this combinatorial problem is
presented below.

First, we consider the open chain with $n$ diamonds. Let us name the total
number of ground states of $n$ diamonds as $w_{n}$. Then, $w_{n}$ can be
expressed as a sum
\begin{equation}
w_{n}=\sum_{k=1}^{n}\left( 3k-1\right) w_{n-k}+\left( 3n+2\right)
\label{rek}
\end{equation}

\begin{figure}[tbp]
\includegraphics[width=5in,angle=0]{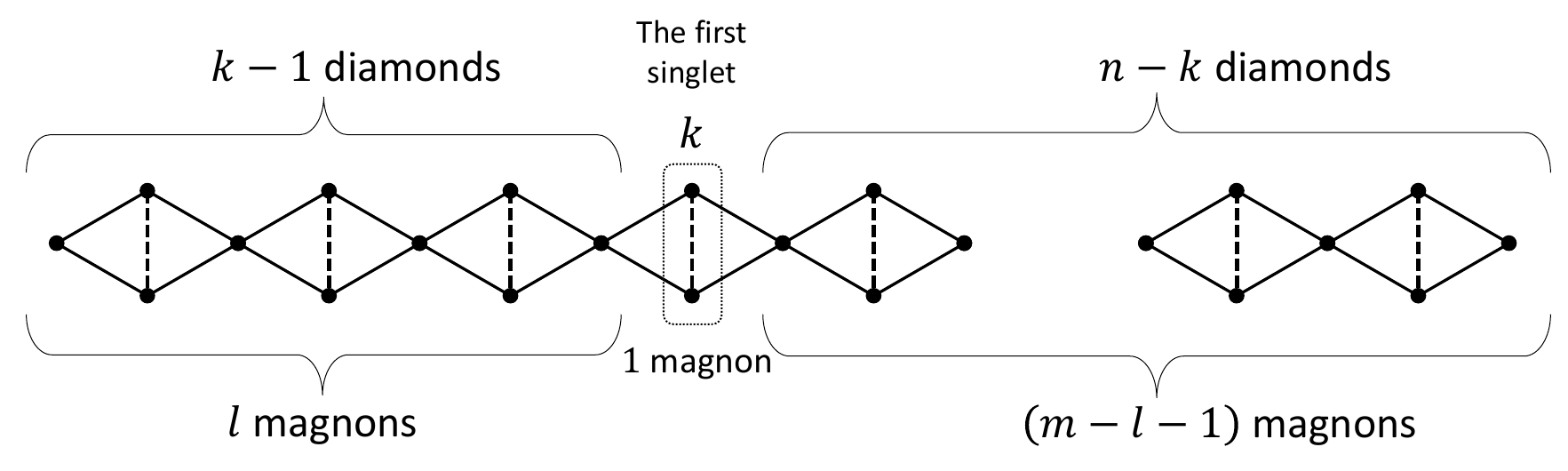}
\caption{Scheme used for the calculation of the ground state
degeneracy for an ideal diamond model.} \label{Fig_open_chain}
\end{figure}

In this expression the first singlet is located in $k$-th diamond,
therefore the remaining part contains $(n-k)$ diamonds with the
total number of states $w_{n-k}$ (see Fig.\ref{Fig_open_chain}).
The factor $(3k-1)$ represents the number of multiplets of the
ferromagnetic state of $(k-1)$ diamonds, containing $(k-1)$
spins-$1$ and $k$ spins-$\frac{1}{2}$. The term with $k=n$ in the
sum corresponds to the special case of just one singlet located on
the last diamond. For this case there is just one
spin-$\frac{1}{2}$ on the right side of the singlet and,
therefore, we put $w_{0}=2$ for this case.

The last term $(3n+2)$ in Eq.(\ref{rek}) corresponds to the case
of no $L=0$ spins in the chain, and the degeneracy of this case is
the number of multiplets of the ferromagnetic state of the total
open chain.

The solution of Eq.(\ref{rek}) is
\begin{equation}
w_{n}=9\cdot 4^{n-1}
\end{equation}

For the calculation of the total number of degenerate states
$W_{n}$ for the cyclic chain, containing $n$ diamonds we write the
equation
\begin{equation}
W_{n}=w_{n-1}+(3n+1)+(n-1)(3n-1)+\sum_{m=1}^{n-2}m\left( 3m+2\right)
w_{n-m-2}  \label{Wn}
\end{equation}

The first term in Eq.(\ref{Wn}) is the number of states when the
last diamond (that loops the chain) contains the diagonal singlet.
The second term $(3n+1)$ in Eq.(\ref{Wn}) corresponds to the case
without singlets in the chain, and the degeneracy in this case is
the number of multiplets of the ferromagnetic state of the total
system. The term $(n-1)(3n-1)$ corresponds to the case with one
singlet on the whole chain. The sum contains terms when the last
diamond belongs to the ferromagnetic section of length $m$.

The sum in (\ref{Wn}) is calculated and gives for $W_{n}$ the
exact result
\begin{equation}
W_{n}=4^{n}+3n-1
\end{equation}

A similar approach can be used for the calculation of the ground
state degeneracy in sectors with fixed total $S^{z}$. Let the
total number of ground states of $n$ diamonds in the sector with
total $S^{z}=S_{\max }^{z}-m $ be $w_{n,m}$. Then $w_{n,m}$ can be
expressed as a sum
\begin{equation}
w_{n,m}=\sum_{k=1}^{n}\sum_{l=0}^{3k-2}w_{n-k,m-l-1}+1  \label{Wnm}
\end{equation}

Here, we follow the method of the calculation of the ground state
degeneracy used in Eq.(\ref{rek}) and shown in
Fig.\ref{Fig_open_chain}. The solutions of Eq.(\ref{Wnm}) are
\begin{eqnarray}
w_{n,0} &=&1 \\
w_{n,1} &=&n+1 \\
w_{n,2} &=&\frac{\left( n+1\right) \left( n+2\right) }{2}
\end{eqnarray}

Generally,%
\begin{equation}
w_{n,k}=W_{n,k}+2W_{n,k-1}+3W_{n,k-2}+2W_{n,k-3}+W_{n,k-4}  \label{Wnk}
\end{equation}%
where $W_{n,k}$ is given in Eq.(\ref{W}) and coincides with the
number of states of $n$ spins $\frac{3}{2}$ in the spin sector
with total $S^{z}=S_{\max }^{z}-m$. Eq.(\ref{Wnk}) means that
effectively the degeneracy of open chain of $n$ diamonds is equal
to that of $(n-1)$ non-interacting spins $\frac{3}{2}$ and two
`edge' spins $1$.

The total number of ground states $W_{n,k}^{PBC}$ of the diamond
chain with PBC in the spin sector with total $S^{z}=S_{\max
}^{z}-k$ is
\begin{equation}
W_{n,k}^{PBC}=W_{n,k}+1 \qquad 0<k\leq \frac{3n}{2}
\end{equation}

All these results are confirmed numerically for finite ideal diamond chains.

Let us consider the point concerning the gap in the excitation
spectrum for the ideal diamond chain. The energy of one-magnon
excitations given by Eq.(\ref{Ek}) for this diamond chain with PBC
is
\begin{equation}
\Delta E_{1}(q)=\frac{3\pm \sqrt{9-8\sin ^{2}\frac{q}{2}}}{2}  \label{dE}
\end{equation}%
where the quasi-momentum is $q=2\pi l/n$. The lowest excited
energy corresponds to $q=2\pi/n$ and is equal to $\Delta
E_1=\frac{2\pi^2}{3n^2}$ at $n\gg 1$.

Now let us consider two-magnon excitations. At first, we consider
the model (\ref{L}) with only one spin $\mathbf{L}=0$ and all
others $\mathbf{L}=1$. As mentioned earlier, the spin
$\mathbf{L}=0$ cuts the chain so that the initially cyclic chain
effectively transforms to the open chain with $n$ spins
$\frac{1}{2}$ (including ends) and $(n-1)$ spins $1$. Let us
consider one-magnon excitations in this open chain, which
correspond to the two-magnon excitations of initial cyclic chain.
The one-magnon spectrum of the open chain is given by
Eq.(\ref{dE}), where $q$ is determined from the equation
\begin{equation}
(1-\sqrt{9-8\sin ^{2}q})\sin (qn)=-2\sin (q(n-2))
\end{equation}

The value of $q$ that corresponds to the lowest excitation,
$\Delta E_{2}$, is $q=\frac{\pi }{2(n-1)}$ at $n\gg 1$. Then
$\Delta E_{2}=\frac{\pi ^{2}}{6n^{2}}$, which is 4x lower than
$\Delta E_{1}$. Numerical calculations show that $\Delta E_{2}$ is
the lowest level in the entire excitation spectrum. Thus, the
spectrum of the ideal diamond chain is gapless in the transition
point and behaves as $\Delta E\sim n^{-2}$.

As well as for the distorted diamond chain at the transition
point, magnetization in the ideal diamond model is zero at $T=0$
and it undergoes a jump from $m=\frac{1}{2}$ in the ferromagnetic
phase to $m=0$ at the transition point. The susceptibility and the
specific heat of the ideal diamond model has been calculated in
\cite{Gu} for a chain with 120 spins. It was shown that the
susceptibility diverges as $\chi \sim \frac{1}{T}$ and $\chi
T=\frac{5}{12}$ at $T\to 0$ as for the distorted diamond chain.
The specific heat $C(T)$ calculated in \cite{Gu} has one maximum.

Deviation from the transition point leads to the ferromagnetic
ground state for $J_{2}<1$ and for $J_{2}>1$ the ground state is
$2^{n}$ degenerate and includes the states with $S_{tot}$ for
$0\leq S_{tot}\leq \frac{n}{2}$. In terms of the Hamiltonian
(\ref{L}) the ground state corresponds to all $L_{i}=0$ and $n$
free spins $s_{i}$. There is a finite gap to the state with
$S_{tot}=\frac{n}{2}+1$, caused by the destruction of one of the
singlets $L_{i}=0$.

\subsection{Diagonal state case}

Let us consider the diamond model for which the the condition
(\ref{c2}) is satisfied but (\ref{c1}) is not (a particular case
is shown in Fig.\ref{Fig_all_cases}d). As it was noted above, the
exact one-magnon states are localized on vertical diagonals of the
diamond chain and they are given by Eq.(\ref{diag}). The exact
ground states in the spin sector $S_{tot}=S_{\max }-k$ $\ (0\leq
k\leq n)$ can be written as a product of $k$ functions
\begin{equation}
\hat{\phi}_{i_{1}}\hat{\phi}_{i_{2}}\ldots \hat{\phi}_{i_{k}}\left\vert
F\right\rangle
\end{equation}%
where the localized states $\hat{\phi}_{i}$ can be located on
neighboring diamonds.

The number of ground states $W(n,k)$ in the spin sector
$S^{z}=S_{\max}-k $ is equal to
\begin{eqnarray}
W(n,k) &=&\sum_{i=0}^{k}C_{n}^{i} \qquad 0\leq k<n \\
W(n,k) &=&2^{n}\qquad n\leq k\leq \frac{3n}{2}
\end{eqnarray}

The total number of ground states $W(n)$ is%
\begin{equation}
W(n)=2^{n}(2n+1) \label{W_trivial}
\end{equation}

The ground state degeneracy in this type of the diamond chain is
exponentially large as well. In accordance with
Eq.(\ref{entropy}), this leads to the residual entropy
$s_{0}=\frac{1}{3}\ln 2$ at $n\gg 1$.

\begin{figure}[tbp]
\includegraphics[width=5in,angle=0]{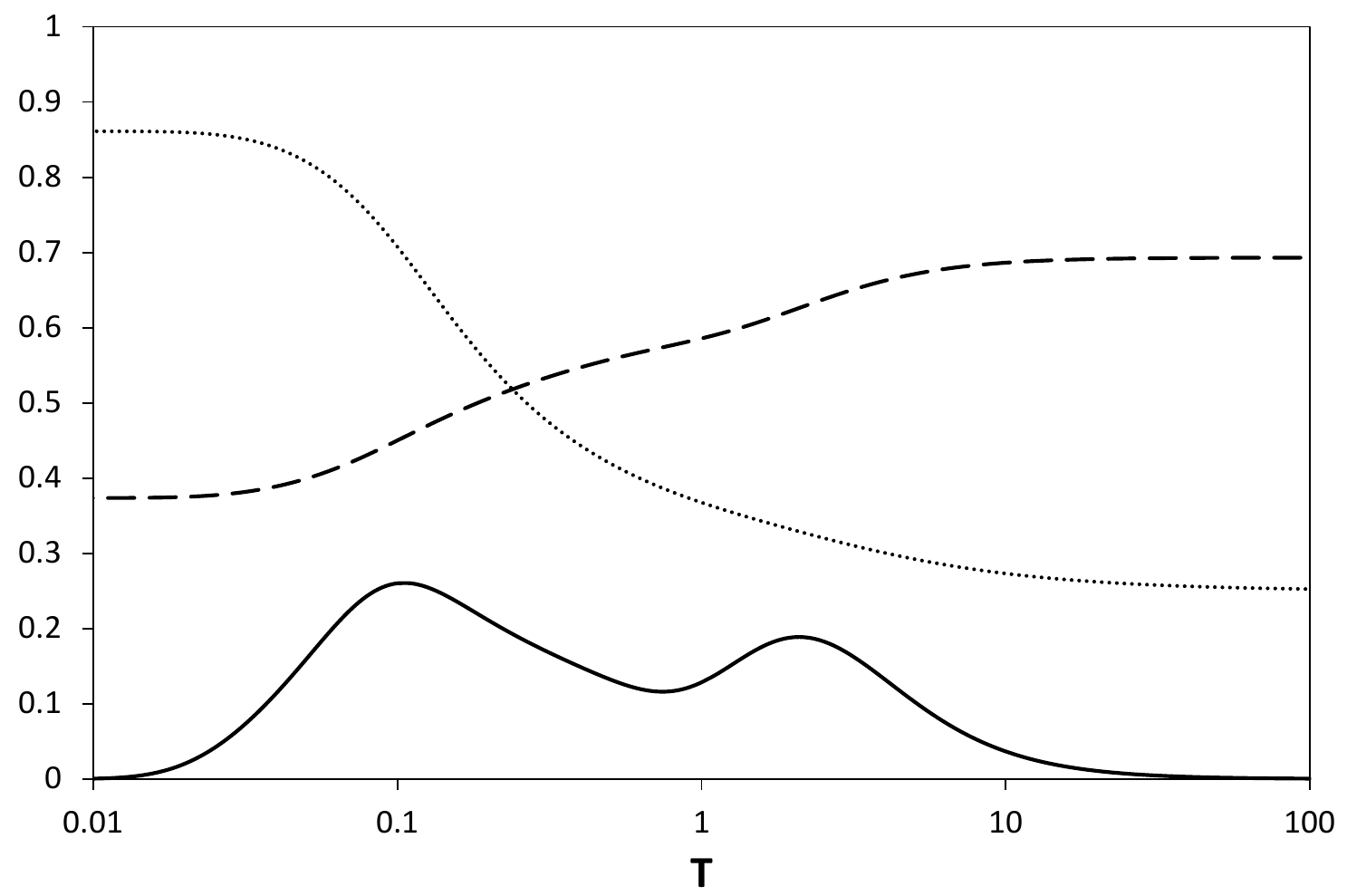}
\caption{Temperature dependencies of the specific heat (solid
line), entropy (dashed line) and the product of magnetic
susceptibility and temperature, $\chi T$, (dotted line) for the
diagonal state case shown in Fig.\ref{Fig_all_cases}d, calculated
numerically for $N=18$. For better visibility the specific heat is
multiplied by a factor of 3, $3C(T)$.} \label{Fig_T_trivial}
\end{figure}

The thermodynamic properties of the diagonal state chain shown in
Fig.\ref{Fig_all_cases}d are demonstrated in
Fig.\ref{Fig_T_trivial}. As it is seen in Fig.\ref{Fig_T_trivial},
the entropy tends to finite value defined by
Eqs.(\ref{entropy}),(\ref{W_trivial}) at low temperatures
$T<0.03$, and approaches to $\ln 2$ at high temperatures. The
specific heat $C(T)$ has two maxima. The magnetic properties of
the diagonal state chain is similar to the general case. The
susceptibility diverges at $T\to 0$ faster than $1/T$. The
magnetization curve in the thermodynamic limit has a form given by
Eq.(\ref{m(h)}). Therefore, the diamond chain of this type is
magnetically ordered at the transition point at $T=0$ and the
magnetization in similarity to the general case of the diamond
chain undergoes a jump from $m=\frac{1}{2}$ in the ferromagnetic
phase to $m=\frac{1}{3}$ at the transition point.

\begin{figure}[tbp]
\includegraphics[width=5in,angle=0]{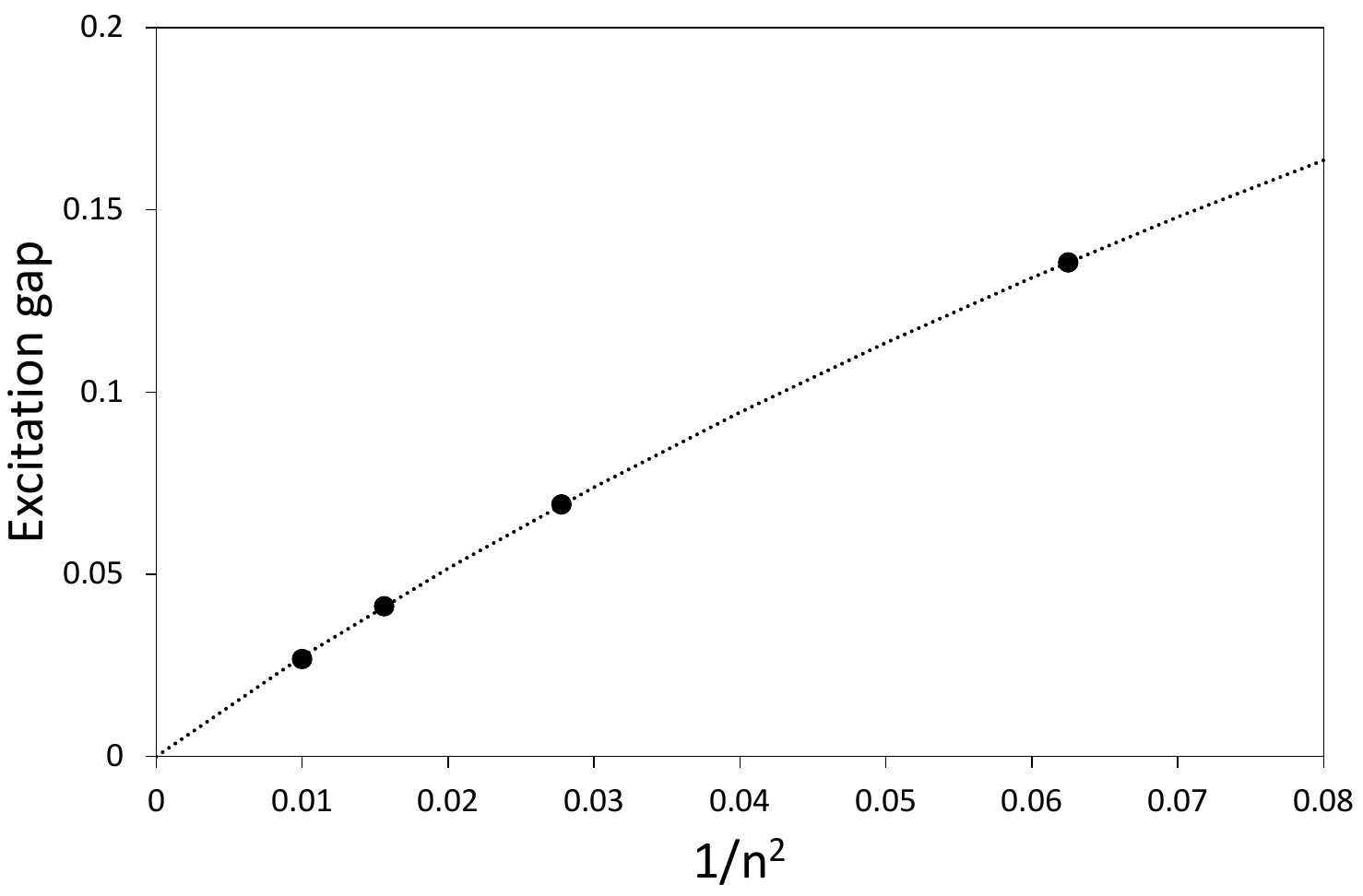}
\caption{Dependence of the energy gap on the system size for
diagonal state case shown in Fig.\ref{Fig_all_cases}d, calculated
numerically for $N=12,18,24,30$ (exact diagonalization and Lanczos
algorithm). Dotted line is a fitting curve $\Delta
E=\frac{3.12}{n^2}-\frac{3.8}{n^3}$.} \label{Fig_gap_trivial}
\end{figure}

The gap in one-magnon excitation spectrum (\ref{Ek}) is equal to
$\Delta E_{1}=\frac{5\pi ^{2}}{6n^{2}}$ for the chain consisting
of diamonds shown in Fig.\ref{Fig_all_cases}d. This means that the
excitation gap goes down as $n^{-2}$ or faster at $n\gg 1$. The
numerical calculations of the gap shown in
Fig.\ref{Fig_gap_trivial} confirm this law, and give $\Delta
E=\frac{3.12}{n^2}$.

\section{Summary}

We have studied the ground state of the spin-$1/2$ Heisenberg
diamond chain with competing F and AF exchange interactions at the
transition point between the ferromagnetic and other (singlet or
ferrimagnetic) ground state phases. The ground state properties of
this model at the transition point are highly nontrivial. The
ground state consists of the localized magnons and special magnon
complexes which form macroscopically degenerate ground state. We
consider four types of diamond chains. One of them is the diamond
chain with only one-magnon localized states. The properties of
this model are similar to those of the F-AF delta-chain. In
particular, the ground state degeneracy is $W(n)\simeq 2^{n}$ and
the residual entropy is $\frac{1}{3}\ln 2$. The diamond chain of
this type is magnetically ordered at $T=0$ and the magnetization
undergoes a jump from $m=\frac{1}{2}$ in the ferromagnetic phase
to $m=\frac{1}{3}$ at the transition point.

The most striking feature of the second considered model, the
distorted diamond chain, is the existence of two- and three-magnon
localized states along with the conventional one-magnon localized
states. These states form flat bands in two- and three magnon
spectrum and they are ground states as well. It turns out that the
number of ground states in the sectors with fixed total spin
projection $S^{z}$ coincides with the one for the system of $n$
non-interacting spins $\frac{3}{2}$. We have checked this fact by
numerical calculations of finite chains. The localized magnon
states in the distorted diamond chain lead to the ground state
degeneracy $W(n)=4^{n}$, which exceeds exponentially the one in
the model with only one-magnon localized states. As a result, the
residual entropy per spin is $s_{0}=\frac{2}{3}\ln 2$. The ideal
diamond chain has the same residual entropy. For the ideal diamond
chain, analytical results have been obtained for the number of the
ground states in different sectors of $S^{z}$ and they are
confirmed by numerical calculations of finite chains. The spectrum
of excitations in the three types of diamond chains is gapless and
it is gapped in the distorted diamond chain. The magnetization is
zero at $T=0$ at the transition point for the distorted and ideal
diamond chains and, therefore, there is a magnetization jump from
$m=\frac{1}{2}$ in the ferromagnetic phase to $m=0$ at the
transition point. The susceptibility per spin $\chi $ diverges as
$\chi \sim 1/T$ and the product $\chi T$ is $\chi T=\frac{5}{12}$
at $T=0$. The main features of four cases of the diamond model are
summarized in the Table.1.

\begin{table}[tbp]
\caption{Properties of four cases of diamond model.}%
\begin{ruledtabular}
\begin{tabular}{ccccc}
& General  & Distorted  & Ideal  & Diagonal  \\
\hline
Ground state degeneracy & $2^{n}+(n+4)C_{n}^{n/2}$ & $4^n$ & $4^n+3n-1$ & $(2n+1)2^n$ \\
Residual entropy per spin & $\frac{1}{3}\ln 2$ & $\frac{2}{3}\ln 2$ & $\frac{2}{3}\ln 2$ & $\frac{1}{3}\ln 2$ \\
Lowest excitation, $\Delta E$  & gapless, $\sim e^{-cn}$ & gapped & gapless, $\sim n^{-2}$ & gapless, $\sim n^{-2}$ \\
Low-T susceptibility, $\chi T$ & diverges & $\frac{5}{12}$ & $\frac{5}{12}$ & diverges \\
Specific heat & two maxima & one maximum & one maximum & two maxima \\
\end{tabular}
\end{ruledtabular}
\end{table}

In this paper we consider the isotropic diamond chain on the
boundary of the ferromagnetic ground state. This model can be
extended to the case of anisotropic interactions similar to those
developed for F-AF delta chain in Ref.\cite{anis,*anis2,*anis3}
(details will be presented elsewhere).

It is known that spin systems with the macroscopic degenerate
ground state show enhanced magnetocaloric effect. From this point
of view the considered diamond chain can serve as a beacon for the
search for materials for low-temperature cooling. Of course,
macroscopic degeneracy occurs for certain relations between
exchange interactions, which can hardly be satisfied exactly in
real compounds. However, some important features of the model
remain if the values of interactions are close to the critical
ones. Recent progress in the synthesis of diamond chain compounds
allows us to expect that compounds with parameters close to those
given in this work can be synthesized in the near future.

\begin{acknowledgments}
The numerical calculations were carried out with use of the ALPS
libraries \cite{alps}. This work was financially supported by the
Ministry of Sciences and Higher Education, Russian Federation
(Research theme state registration number 122041400110-4).
\end{acknowledgments}

\bibliography{diamond}

\end{document}